\documentclass[acmsmall]{acmart}
\usepackage{wrapfig}
\usepackage{amsfonts}
\usepackage{ulem}

\AtBeginDocument{%
  \providecommand\BibTeX{{%
    \normalfont B\kern-0.5em{\scshape i\kern-0.25em b}\kern-0.8em\TeX}}}

\setcopyright{acmcopyright}
\copyrightyear{2021}
\acmYear{2021}
\acmDOI{10.1145/1122445.1122456}

\acmJournal{JETC}
\acmVolume{0}
\acmNumber{0}
\acmArticle{0}
\acmMonth{0}

\acmSubmissionID{JETC-2020-0271.R1}

\begin{document}

\title{RT-RCG: Neural Network and Accelerator Search Towards Effective and Real-time ECG Reconstruction from Intracardiac Electrograms}

{\let\thefootnote\relax\footnote{{This work was supported in part by the 
National Institutes of Health under Grant R01HL144683 and, 
National Science Foundation under Grant CCF-1838873.}}}
\author{Yongan Zhang}
\author{Anton Banta}
\author{Yonggan Fu}
\affiliation{%
  \institution{Rice University}
  \streetaddress{6100 Main ST}
  \city{Houston}
  \country{USA}}
\email{{yz87; arb17; yf22}@rice.edu}




\author{Mathews M. John}
\author{Allison Post}
\author{Mehdi Razavi}
\affiliation{%
  \institution{Texas Heart Institute}
  \streetaddress{6770 Bertner Ave}
  \city{Houston}
  \country{USA}}
\email{{mjohn; apost}@texasheart.org, mehdirazavi1@gmail.com }


\author{Joseph Cavallaro}
\author{Behnaam Aazhang} 
\author{Yingyan Lin}
\affiliation{%
  \institution{Rice University}
  \streetaddress{6100 Main ST}
  \city{Houston}
  \country{USA}}
\email{{cavallar; aaz; yingyan.lin}@rice.edu}




\renewcommand{\shortauthors}{Zhang, et al.}

\begin{abstract}

There exists a gap in terms of the signals provided by pacemakers (i.e., intracardiac electrogram (EGM)) and the signals doctors use (i.e., 12-lead electrocardiogram (ECG)) to diagnose abnormal rhythms. Therefore, the former, even if remotely transmitted, are not sufficient for doctors to provide a precise diagnosis, let alone make a timely intervention. To close this gap and make a heuristic step towards real-time critical intervention in instant response to irregular and infrequent ventricular rhythms, we propose a new framework dubbed RT-RCG to automatically search for (1) efficient Deep Neural Network (DNN) structures and then (2) corresponding accelerators, to enable \textbf{R}eal-\textbf{T}ime and high-quality \textbf{R}econstruction of E\textbf{C}G signals from E\textbf{G}M signals. Specifically, RT-RCG proposes a new DNN search space tailored for ECG reconstruction from EGM signals, and incorporates a differentiable acceleration search (DAS) engine to efficiently navigate over the large and discrete accelerator design space to generate optimized accelerators. Extensive experiments and ablation studies under various settings consistently validate the effectiveness of our RT-RCG. To the best of our knowledge, RT-RCG is the first to leverage neural architecture search (NAS) to simultaneously tackle both reconstruction efficacy and efficiency.

\end{abstract}



\maketitle

\section{Introduction}
 \label{sec:intro}

Over 5.8 million people in the USA and over 23 million worldwide are affected by cardiac diseases ~\cite{heron2016changes,bui2011epidemiology}, where the inability to generate or conduct the electrical signals necessary to stimulate muscle contraction is the major cause for many heart failures~\cite{chen2018}. To treat these failures, artificial electronic pacemakers are usually implanted to stimulate the heart with electrical impulses to maintain or restore a normal rhythm. In particular, about 3 million people worldwide use pacemakers and  6,000,000 pacemakers are implanted each year~\cite{wood2002cardiac}.
Currently, cardiac pacemakers cannot sense or compute 12-lead ECGs from EGMs. Patients with pacemakers require regular, and often costly and time-consuming hospital visits to ensure (1) the proper functioning of the pacemaker and (2) the timely adjustment of the pacing parameters to adapt to changes in the heart's condition over time. Thanks to recent advances in the internet of things (IoT) technologies, remote monitoring of pacemakers will become more commonplace, allowing doctors to check  pacemaker status and thus reducing the frequency of costly hospital visits~\cite{yeole2016use}.

Despite the promising advantages of remotely monitoring pacemakers, there is still a gap in terms of the signals that can be provided by pacemakers and the ones doctors need 
to diagnose abnormal rhythms and provide appropriate therapy.
Specifically, cardiac pacemakers utilize continuously collected EGMs which are electrical activities sensed locally via  implanted electrodes. However, 12-lead ECGs obtained from skin electrodes contain significantly greater information than EGMs which, in certain cases, could be utilized to better diagnose abnormal rhythms and provide appropriate therapy. 
To close this gap, the synthesis or reconstruction of ECG signals from a set of EGM signals is of great significance in enabling effective remote monitoring of pacemakers, providing necessary therapy, and making timely clinical intervention possible~\cite{van2014value}. As such, there has been a growing interest in developing techniques to reconstruct ECG signals from their corresponding EGM signals using linear filtering \cite{gentil2005surface, kachenoura2007surface, kachenoura2008using, mendenhall2010implantable}, fixed dipole modeling algorithms \cite{mendenhall201012, mendenhall2010implantable}, and nonlinear reconstruction via a time delay neural network \cite{kachenoura2009non, kachenoura2009comparison, poree2012surface}. 

While the aforementioned techniques were pioneering steps, there is still much room to improve their performance for practical and widespread adoption. In particular, most of the existing techniques adopt either linear approaches that lack generalization capability for unseen symptoms and can fail in the presence of noises and artifacts, or a multivariate nonlinear approach that requires the simultaneous recording of both EGM and 12-lead ECG signals for every single patient \cite{kachenoura2009non, kachenoura2009comparison, poree2012surface}. Motivated by the recent breakthroughs in deep neural networks (DNNs) and their demonstrated promise in medical applications \cite{eraslan2019single,erhan2010does,tran2017missing, cosentino2020provable}, it is natural to consider DNN based reconstruction techniques, aiming for much improved generalization capability and better efficacy towards more practical clinical uses. However, the excellent performance of DNN based solutions often comes at the cost of high complexity (e.g., millions of parameters and operations \cite{wu2018deep}) which stands at odds with the extremely constrained resources at the implanted, battery-powered pacemakers. Specifically,
 restricted by the pacemaker's limited hardware budget, the often complex DNN based solutions make it particularly challenging to handle real-time reconstruction on the pacemakers,
which could enable improved and possibly life-critical interventions to the patients. Currently, EGMs stored in pacemakers are analyzed offline through an inpatient setting for improved diagnosis of the underlying condition, where therapeutic intervention might need to be changed over time and thus require real-time adaptation. For example, monitoring ECG data in real-time can allow for determination of potentially deadly ventricular arrhythmias~\cite{nof2013catheter}, and dictate pacing mediated therapies such as anti-tachycardia pacing. Online real-time reconstruction of EGMs to ECGs allows for real-time and immediate intervention and thus potentially paves the way for novel treatments, whereas offline reconstruction may not always be possible and the potential latency involved in doing so could be life threatening. Another example is utilizing ECGs in real-time for optimizing parameters for cardiac resynchronization therapy to treat heart failure patients~\cite{antonini2012optimization}, where a real-time embedded accelerator allows for on-device reconstruction with a low latency and is thus critical. Furthermore, with traditional pacemakers slowly being replaced by leadless pacemakers~\cite{tjong2017permanent}, such an accelerator would also pave the way for improved therapy with minimal sensing sites.

To this end, we aim to develop an efficient DNN based reconstruction framework to push forward the efficacy and efficiency frontier towards practical and widespread adoption by leveraging recent advances in neural architecture search and DNN acceleration. Specifically, we make the following contributions in this work: 

\begin{itemize}
\vspace{-0.5em}
\item We propose a new framework dubbed RT-RCG, which can automatically search for (1) efficient DNN structures and then (2) corresponding accelerators to enable \textbf{R}eal-\textbf{T}ime and high-quality \textbf{R}econstruction of E\textbf{C}G signals from E\textbf{G}M signals.
To the best of our knowledge, the proposed RT-RCG is the first to simultaneously tackle and leverage neural architecture search (NAS) for both reconstruction efficacy and efficiency.    

\item Drawing inspiration from existing ECG reconstruction works, RT-RCG proposes a new DNN search space tailored for ECG reconstruction from EGM signals to enable automated search for DNNs which consistently outperform state-of-the-art (SOTA) reconstruction techniques in terms of both reconstruction correlation (between the reconstructed ECGs and the real-measured ECGs) and algorithmic generalization capability. 

\item Built upon recent advances in DNN acceleration, RT-RCG incorporates a differentiable acceleration search (DAS) engine which makes use of gradient-based optimization to efficiently navigate over the large and discrete accelerator design space to automatically generate optimized accelerators that achieve real-time reconstruction. 

\item Extensive experiments and ablation studies under various settings consistently validate the effectiveness of our proposed RT-RCG in leading to higher reconstruction quality and better reconstruction efficiency as compared to SOTA reconstruction algorithms and DNN accelerators, respectively. We believe that RT-RCG has made a nontrivial step towards practical ECG reconstruction from EGM signals on the pacemaker, promising the real possibility of real-time critical intervention in instant response to irregular and infrequent ventricular rhythms that require timely treatment.

\end{itemize}

\section{Related works}
\textbf{ECG Reconstruction.}
In response to the practical need of ECG reconstruction from EGM signals, various methods have been proposed~\cite{gentil2005surface,kachenoura2007surface,kachenoura2008using,kachenoura2009comparison,kachenoura2009non,mendenhall201012,mendenhall2010implantable,poree2012surface} using linear filtering~\citep{gentil2005surface,kachenoura2007surface,kachenoura2008using,mendenhall201012}, fixed dipole
modeling algorithms~\citep{mendenhall2010implantable}, nonlinear filtering~\cite{kachenoura2009comparison,kachenoura2009non}, and time delay neural networks~\cite{poree2012surface},
In particular, a single EGM channel was used to synthesize a single ECG lead in \cite{gentil2005surface}, which can be highly dependent on the chosen EGM lead; Later, logical extension of \cite{gentil2005surface} were developed which uses all EGM leads for synthesis \cite{kachenoura2007surface, kachenoura2008using}, where both the EGMs and the ECGs were first projected onto a 3D space and then three linear filters were calculated between the signals, providing an indirect way to find the transfer functions between EGM signals and the 12-lead ECG; Similarly, \cite{mendenhall201012, mendenhall2010implantable} directly calculated a multivariate linear transfer matrix between the EGMs and the 12-lead ECGs via penalized linear regression. Despite their satisfactory performance, especially for patients with a surface ECG containing only a one beat morphology, these linear methods can suffer from a degraded correlation between the EGMs and the ECGs in real applications due to the noises and artifacts present, and the natural evolution and diversity of the pathology. 
The limitation of linear reconstruction methods (e.g., an average correlation value of lower than 0.5 in \cite{mendenhall201012}) motivated the multivariate nonlinear approach presented in \cite{kachenoura2009non, kachenoura2009comparison, poree2012surface}, all of which require the simultaneous recording of the EGMs and 12-lead ECGs for every single patient to train a time-delay artificial neural network (TDNN). While this method provided the best average correlation results for sinus rhythm heartbeats, it is still limited for practical uses as it  cannot effectively reconstruct diseased morphologies and $12$ different TDNN models must be calculated to reconstruct each ECG lead.

Built upon the above prior works, RT-RCG targets reconstruction algorithms that are generally applicable in the presence of noise, artifacts, and diverse pathologies. 


\textbf{DNNs in Cardiology Applications.}
The recent breakthroughs of DNNs in various fields have sparked a growing interest in developing DNN based solutions for cardiologic problems spanning from ECG classification to sleep status monitoring~\cite{xiong2016ecg,li2018method,elola2019deep,xu2018towards,hannun2019cardiologist,hanbay2018deep}. In particular, \cite{xiong2016ecg} adopted a DNN to remove noises contaminating the ECG signals; ~\cite{elola2019deep} used two DNNs together with short-duration (5 seconds) ECG segments to detect pulses during out-of-hospital cardiac arrest;~\cite{xu2018towards} proposed to utilize DNNs for the classification of ECG signals into different heart rhythms (i.e., normal beat or different types of arrhythmias);~\cite{li2018method} made use of a DNN and a hidden Markov model to detect obstructive sleep apnea based on single lead ECG signals. The readers are referred to ~\cite{bizopoulos2018deep} for a detailed survey on applying DNNs to cardiology applications. While these works demonstrate the great potential of DNN based solutions for cardiologic problems, DNN-powered ECG-EGM reconstruction algorithms are still under-explored, let alone real-time reconstruction implementation, motivating us to propose and develop our RT-RCG framework.

\textbf{Neural Architecture Search.}
Neural architecture search (NAS)~\cite{zoph2016neural} has emerged as one of the most significant sub-fields of AutoML \cite{hutter2019automated} as it enables automatically searching for an optimal DNN structure from the given data and has outperformed manually designed DNNs on a range of tasks such as image classification~\cite{tan2019efficientnet, tan2019mnasnet, howard2019searching, liu2018darts} and segmentation~\cite{chen2018searching, liu2019auto, chen2019fasterseg}. Early NAS works achieve SOTA performance at the cost of enormous search time~\cite{zoph2016neural, zoph2018learning, real2019regularized}. Specifically, reinforcement learning (RL) based NAS~\cite{zoph2016neural, zoph2018learning, tan2019mnasnet, howard2019searching, tan2019efficientnet} and evolutionary algorithm based NAS~\cite{pham2018efficient, real2019regularized} explored the search space and train each sampled network candidate from scratch, thus suffering from prohibitive search costs. 
Later, differentiable NAS (DNAS)~\cite{liu2018darts, wu2019fbnet, wan2020fbnetv2, cai2018proxylessnas, xie2018snas} was proposed to update the weights and architecture in a differentiable manner through supernet weight sharing, reducing the search time to several hours~\cite{stamoulis2019single}. Motivated by the promising performance achieved by those DNAS works, recent works have extended DNAS to more tasks such as segmentation~\cite{liu2019auto, chen2019fasterseg}, image enhancement~\cite{fu2020autogan,lee2020journey}, and language modeling~\cite{chen2020adabert}. 
As a result, we leverage the DNAS method integrated with a new search space to develop our
proposed RT-RCG framework.

\textbf{DNN Accelerators.}
\label{sec:related_dnn_accelerators}
DNNs' powerful performance comes at a cost of a prohibitive complexity, motivating extensive research in dedicated DNN accelerators as specialized hardware has the potential to achieve orders-of-magnitude higher energy/time efficiency. Specifically, it has been shown that aggressive efficiency can be achieved by carefully designing the micro-architectures (e.g., the number of memory hierarchies or processing element (PE) units, the storage size of different memories, and the shape of the PE array) and algorithm-to-hardware mapping strategies (i.e., dataflow). For example, representative works, such as ShiDiannao \cite{du2015shidiannao} and Eyeriss \cite{chen2017eyeriss}, identified the performance bottleneck caused by the required massive data movements and proposed novel micro-architectures and dataflows that aim to maximize data reuse for reducing the energy/time cost to access higher cost memories. 
Early works mostly rely on experts' manual design, which can be very time-consuming (months or even years) and require cross-disciplinary knowledge in algorithm, micro-architecture, and circuit design. In response to the intense demands and challenges of manually designing DNN accelerators, we have seen rapid development of design flow~\cite{vivado_HLS,hls_chen2005xpilot,hls_chen2009lopass,hls_rupnow2011high} and DNN design automation frameworks~\cite{wang2016deepburning, zhang2018caffeine, guan2017fp, venkatesanmagnet,wang2018design} to standardize the design flow of DNN accelerators and to expedite the development process. For example, the DNNBuilder accelerator~\cite{zhang2018dnnbuilder} applied an automated resource allocation strategy, fine-grained layer-based pipeline, and column-based cache to deliver high-quality FPGA-based DNN accelerators, and \cite{xu2020autodnnchip} made the first step towards automatically generating both FPGA- and ASIC-based DNN accelerators without humans in the loop given the DNNs from machine learning frameworks (e.g., PyTorch) for a designated application and dataset.

Leveraging the learning from prior works, RT-RCG integrates an DAS engine to automatically generate micro-architectures and dataflows to achieve real-time reconstruction.

\textbf{DNN Algorithm and Accelerator Co-exploration.} 
Exploring the networks and the corresponding accelerators in a joint manner~\cite{edge_fpga_co_design, abdelfattah2020best, yang2020co, jiang2020device, jiang2019hardware, li2020edd} has shown great potential towards maximizing both accuracy and efficiency. 
Recent works have extended NAS to jointly search DNN accelerators in addition to DNN structures. In particular,~\cite{edge_fpga_co_design, abdelfattah2020best,jiang2019hardware,yang2020co} conducted RL-based searches to co-explore the network structures and design parameters of an FPGA-/ASIC-based accelerator, but their RL-based methods can suffer from large search costs, limiting their scalability to handle large joint spaces. Recently, \cite{li2020edd,choi2020dance} extended differentiable NAS to network and accelerator co-search. However,~\cite{li2020edd} only considered one accelerator parameter (i.e., the parallel factor of an FPGA accelerator template) which is not always applicable to most naturally non-differentiable accelerator design parameteres such as loop order and loop size, while~\cite{choi2020dance} adopted a DNN to generate accelerator designs with network structures as the DNN's inputs, which lack interpretability. In contrast, our work adopts differentiable joint search in a sequential manner to efficiently explore a generic network and accelerator design space.

\vspace{-0.5em}
\section{Preliminaries of Deep Neural Networks (DNNs) and the EGM/ECG Data Format}
\label{sec:preliminary}

\begin{figure}
\vspace{-1.2em}
\includegraphics[width=0.8\columnwidth]{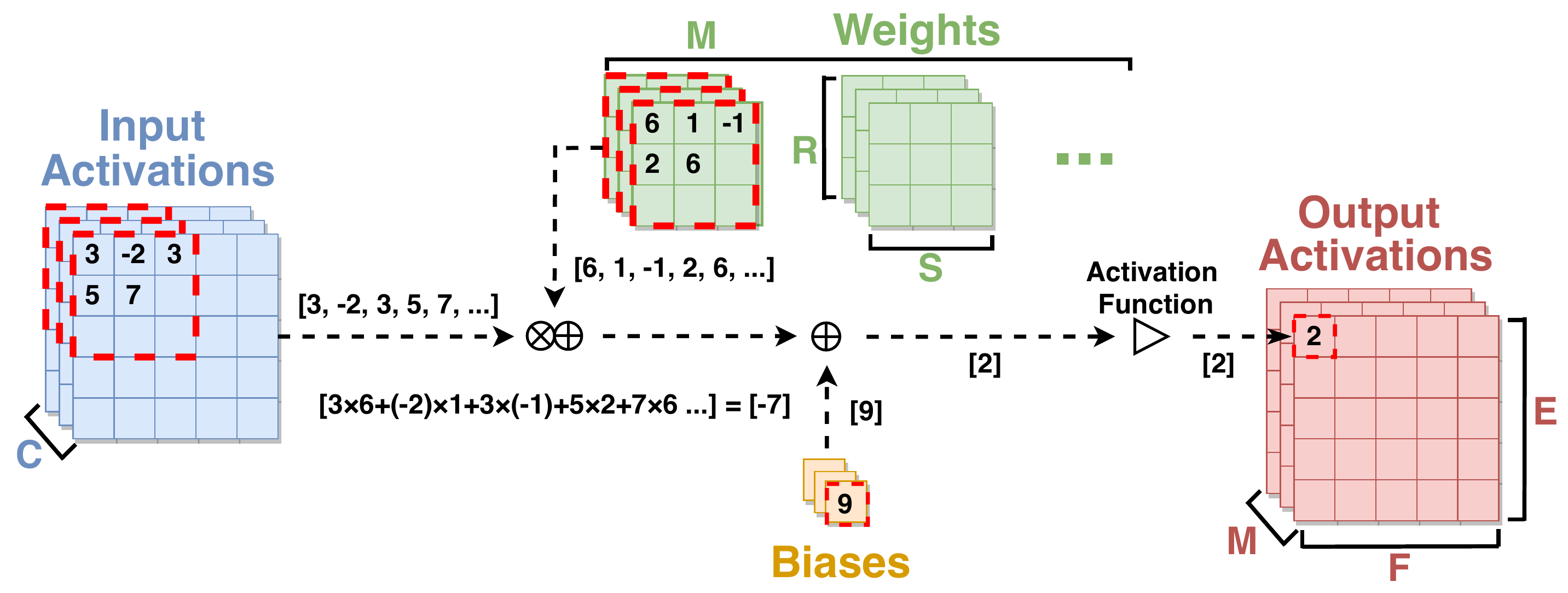}
\vspace{-1.5em}
\caption{An illustrative example of one CONV operation as formulated in Equation~(\ref{eq:CONV}), where $M$ / $C$ (the number of input / output channels), $E$ / $F$ (the input feature map height / width), $R$ / $S$ (kernel 
height / width) and U (stride) are 3 / 3, 5 / 5, 3 / 3, and 1, respectively. This example assumes that ReLU is used as the activation function and the first output is 2. }
\vspace{-1.6em}
\label{fig:example_conv}
\end{figure}

\textbf{Deep Neural Networks (DNNs).}
Modern DNNs usually consist of a cascade of multiple convolutional (CONV), pooling, and fully-connected (FC) layers through which the inputs are progressively processed. The CONV and FC layers can be described as:

\begin{equation}
    \begin{split}
    \textit{\textbf{O}}[c_o][e][f]=\sigma((\sum_{c_i=0}^{C-1}&{\sum_{k_r=0}^{R-1}{\sum_{k_s=0}^{S-1}{\textit{\textbf{W}}[c_o][c_i][k_r][k_s]}\times \textit{\textbf{I}}[c_i][eU+k_r][fU+k_s]}})+\textbf{B}[c_o])\\
    & 0\le c_o < M,~0\le e < E, 0\le f < F 
    \end{split}
    \label{eq:CONV}
\end{equation}
where \textit{$\textbf{W}$}, \textit{$\textbf{I}$}, \textit{$\textbf{O}$}, and \textit{$\textbf{B}$} denote the weights, input activations, output activations, and biases, respectively. 
In the CONV layers (see an example in Figure~\ref{fig:example_conv}), $C$ and $M$, $E$ and $F$, $R$ and $S$, and $U$ stand for the number of input and output channels, the size of input and output feature maps, and the size of weight filters, and stride, respectively; while in the FC layers, $C$ and $M$ represent the number of input and output neurons, respectively; with $\sigma$ denoting the activation function, e.g., a $ReLU$ function ($ReLU(x)=max(x,0)$). The pooling layers reduce the dimension of feature maps via average or max pooling. The recently emerging compact DNNs (e.g., MobileNet~\cite{howard2017mobilenets} and EfficientNet~\cite{tan2019efficientnet}) introduce depth-wise CONV layers and squeeze-and-excite layers which can be expressed in the above description as well~\cite{chen2019eyeriss}. 

\textbf{Pre-processing of the EGM/ECG Signals.} Here we describe the adopted pre-processing for the EGM and ECG signals, both of which were recorded simultaneously during the cardiac ablation procedure.
In the first step, the signals were initially obtained at a sampling frequency of 1000 Hz, and subsequently bandpass filtered using a 5-th order Butterworth filter with a cutoff frequency at $3$ Hz and $50$ Hz. The cutoff at 3 Hz helps to eliminate potential baseline wanders and the cutoff at 50 Hz can eliminate powerline interferences, electromyographic noise, and electrode motion artifact noise \cite{kher2019signal}.
To align with the phase change caused by the pre-processing filtering in the forward direction, we adopt the zero phase filtering and also filter the signal backwards in time \cite{kormylo1974two} to ensure that the pre-processing of the data does not introduce additional distortion.
In the second step of the pre-processing, the data from the previous step is segmented to extract the QRS portion of ECG signals which contains much information about the synchronization of the heart's ventricles and has been demonstrated to be a strong biomarker for overall cardiac health \cite{moulton1990premature}.

\textbf{Time-frequency Representation of EGM/ECG Signals.} To make use of DNNs to reconstruct ECG signals from EGM signals, we first transfer EGM signals' $2$-dimensional (2D) multi-channel time-series representation into a  $3$-dimensional (3D) time-frequency representation with the help of STFT, inspired by the similar treatments in speech recognition and audio processing applications \cite{amodei2016deep, parveen2004speech, xu2014regression}. 

Assuming that the matrix $S_t \in \mathbb{R}^{M \times T}$ denotes the EGM time-series where $M$ and $T$ correspond to the number of channels and the number of time samples for each channel, respectively, then $S_t$ can be re-formulated as $S_t = \left [\textbf{s}_t^{(1)}, \dots, \textbf{s}_t^{(m)}, \dots, \textbf{s}_t^{(M)} \right ]^{\top}$, with $\textbf{s}_t^{(m)} \in \mathbb{R}^{T \times 1}$ and ${\top}$ denoting the time-series for each of the $M$ channels and the transpose operator, respectively. As such, the corresponding 3D time-frequency signals, denoted as $S_{tf}$, can be represented as:

\begin{align} 
    \begin{split}
    S_{tf}=&\left [\textbf{s}_{tf}^{(1)},\dots, \textbf{s}_{tf}^{(M)} \right ]^{\top}\\
    \textbf{s}_{tf}^{(m)} = & \left [  \textbf{s}_t^{(m)} \circ \textbf{h}_{1},\dots, \textbf{s}_t^{(m)} \circ \textbf{h}_{K}                \right ]^{\top}
    \end{split}
    \label{eq:data_format}
\end{align}
where $\forall k \in \left \{1,\dots,K \right \}$, $\textbf{h}_k \in \mathbb{C}^{T}$ defines the $k$-th time-frequency filter in the complex space corresponding to $\textbf{s}_t^{(m)}$, and $\circ$ denotes the convolution operator. 
We set the length of each time-frequency filter (after filtering) as $T_f$, and thus $\textbf{s}_{tf}^{(m)} \in \mathbb{C}^{K \times T_f}$ represents a 2D complex matrix with each row denoting the time domain information and each column denoting the frequency domain information.
Concatenating all channels' time-frequency representation, we then have a 3D complex matrix $S_{tf}\in \mathbb{C}^{M \times K \times T_f}$. In this work, we use windowed Fourier filters as the filters $\textbf{h}_k$, i.e., transferring the time-series representation into its time-frequency one which becomes the operation of applying a 3D short-time Fourier transform (STFT) operator to the time-series EGM signals.

\vspace{-0.5em}
\section{The Proposed RT-RCG Framework}
 \label{sec:method}


\subsection{RT-RCG: Overview and Problem Formulation}
 \label{sec:overview_formulation}
\begin{figure}[t!]
  \vspace{-2em}
    \centering
    \includegraphics[width=0.85\linewidth]{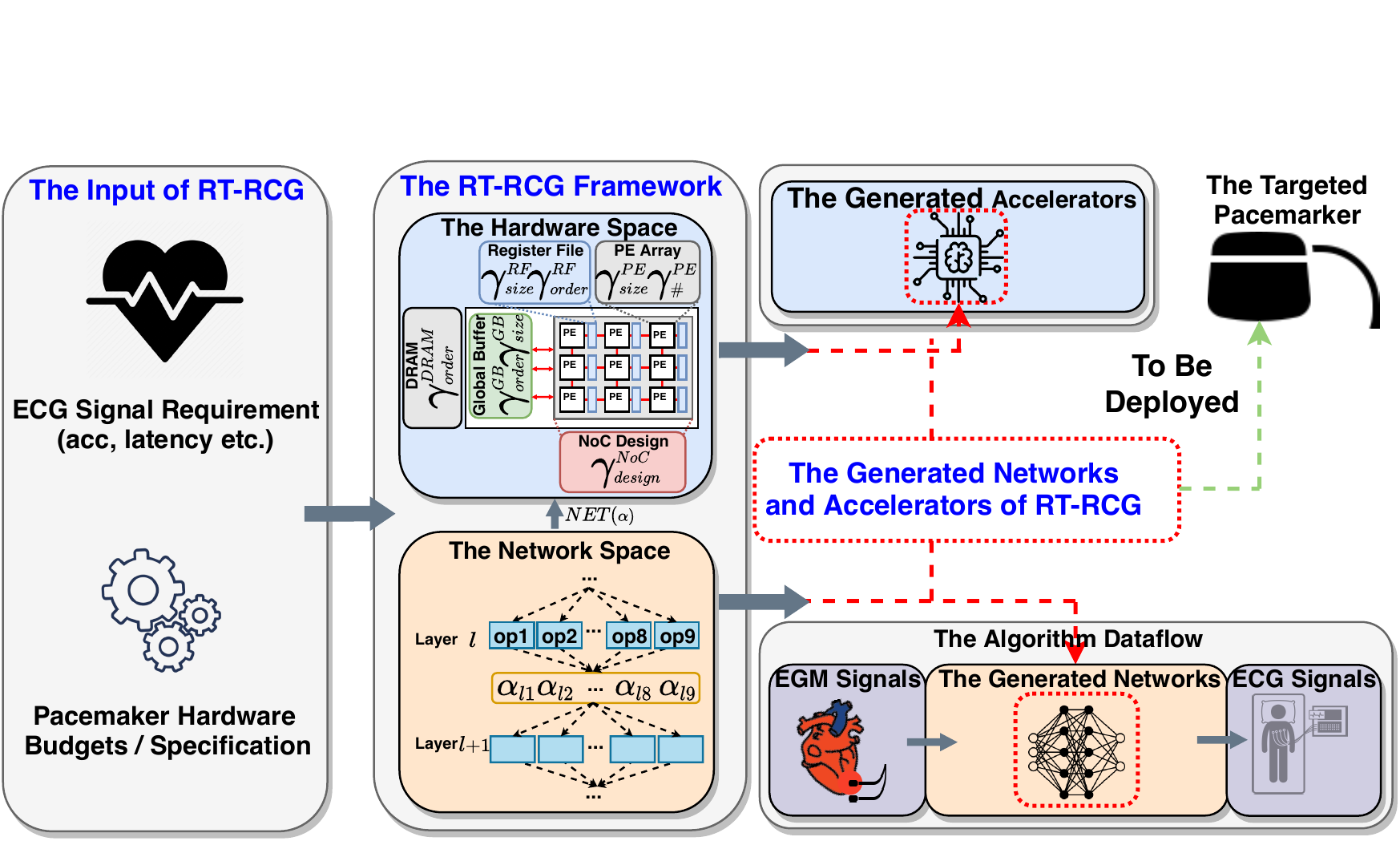}
     \vspace{-1em}
         \caption{An overview of the proposed RT-RCG framework, which accepts the recorded EGM/ECG signals dataset and the target hardware specification as inputs to automatically generate reconstruction networks and their corresponding accelerators to maximize the reconstruction quality and acceleration efficiency.
    }
    \label{fig:co-search}
    \vspace{-2em}
\end{figure}

\vspace{-0.5em}
\textbf{Framework Overview.}
Figure~\ref{fig:co-search} shows an overview of the proposed RT-RCG framework. 
Given the recorded EGM signals, user-specified demands (e.g., accuracy and latency), and hardware resource budgets/specification, our RT-RCG framework automatically searches for networks to maximize the reconstruction efficacy and then the corresponding accelerators to maximize the hardware acceleration efficiency, i.e., the outputs of RT-RCG include (1) the searched network to be used for reconstructing ECG signals from the input EGM signals and (2) the searched accelerator to process the searched network with optimized hardware efficiency. 
In particular, our RT-RCG framework consists of two components, i.e., a differentiable network search (DNS) engine and a DAS engine which will be described in Section \ref{sec:DNS} and Section \ref{sec:DAS}, respectively.

\textbf{The Optimization Formulation.} As stated in Section \ref{sec:intro}, RT-RCG is designed to reconstruct the full 12-lead ECG signals from the recorded (partial) EGM signals, with both originally being time-series signals. For notation, we denote the EGM and ECG samples using $\left \{ X_{n_t} \right \}_{n=1}^{N}$ and $\left \{ Y_{n_t} \right \}_{n=1}^{N}$, respectively, where $N$ denotes the total number of heartbeats in the dataset (see Table~\ref{tab1}). Meanwhile, the EGM and ECG signals can be represented using a 2D matrix, i.e., $X_{n_t} \in \mathbb{R}^{M_{EGM} \times T}$ and $Y_{n_t} \in \mathbb{R}^{M_{ECG} \times T}$, where $M_{EGM}$ and $M_{ECG}$ denote the number of channels (leads) for the ECG and EGM signals, respectively, and T denotes the number of time samples per heartbeat.
As introduced in Section~\ref{sec:preliminary}, the ECG and EGM signals will first be transferred into a time-frequency format denoted as $X_{n_{tf}} \in \mathbb{R}^{M_{EGM} \times K\times T_f}$ and $Y_{n_{tf}} \in \mathbb{R}^{M_{ECG} \times K\times T_f}$, respectively. In this work, we have $M_{EGM}=5$ and $M_{ECG}=12$, respectively, and both $K$ and $T_f$ are empirically fixed to 16 with a STFT window size of 30 and overlap of 6 during the filtering, based on the collected dataset (see Table~\ref{tab1}).
Through empirical studies, this STFT configuration gave us the best subsequent reconstruction accuracy with the least number of parameters. As such, the problem of reconstructing ECG from EGM becomes how to map the signals in 
$\mathbb{R}^{M_{EGM} \times K \times T_f}$ to that in $\mathbb{R}^{M_{ECG}\times K \times T_f}$,
which can be considered as a problem of multivariate regression and the corresponding optimization can be formulated as follows: 

\begin{equation}
    \min\limits_{f \in \mathcal{H}}\sum\limits_{n=1}^N{\mathcal{L}(f(X_{n_{tf}}),Y_{n_{tf}})}
    \label{eq:loss} 
\end{equation}
\noindent where $\mathcal{H}$ denotes the function space, $f$ denotes the reconstruction function that aims to reconstruct $Y_{n_{tf}}$ given $X_{n_{tf}}$, $\mathcal{L}$ denotes the loss function of reconstruction capturing the total difference (e.g., the mean square error) between the reconstructed samples $f(X_{n_{tf}})$ and the real-measured samples $Y_{n_{tf}}$ for all the ${N}$ samples. The goal of the optimization is to find a reconstruction function $f$ that minimizes the reconstruction loss $\mathcal{L}$. In RT-RCG, we use a DNN to approximate and search for $f$ using RT-RCG's DNS engine, with  
the direct output of $f$ having a time-frequency format and then being transferred back into a time-series format for evaluating the reconstruction efficacy. 
During training, the negative Pearson correlation~\cite{benesty2009pearson} of the flattened time-frequency data between the reconstructed ECG and the corresponding real-measured ECG signals will be used as the loss for optimization. For evaluation, the Pearson correlation will be calculated between the reconstructed and corresponding real-measured ECG signals on a test set (excluded in training) after both of them are converted back to the time domain through the inverse STFT. 
Note that the (inverse) STFT process will neither be accelerated by the proposed RT-RCG's hardware nor be counted towards the final latency in our experiments. This is because for a single piece of input, the combined operations of both STFT and inverse STFT only take up about 1\% of the total operations in the inference when DNNs shown in Table~\ref{tab:15-layer} are considered, assuming a fast convolution algorithm is adopted. The (inverse) STFT operation can thus be easily conducted on the hardware accelerator's accompanying CPU incurring a negligible latency overhead.

\vspace{-0.5em}
\subsection{RT-RCG: The DNS Engine }
\vspace{-0.3em}
\label{sec:DNS} 
\textbf{The Network Search Space.}
\begin{table} [!b]
\vspace{-1.8em}
\centering
\caption{Visualizing RT-RCG's network backbone with 14 searchable blocks, where TBS denotes "To Be Searched".}\label{tab:backbone}\vspace{-1em}
\scriptsize
\resizebox{\textwidth}{!}{
    \begin{tabular}{cccccccccccccccc}\toprule
    Operation type & CONV & Maxpool & CONV & Maxpool& \textbf{Searchable blocks $\times$ 14} & DECONV & Upsample & DECONV & Upsample& CONV & CONV& CONV\\\midrule \midrule
    Output channels &48 & - &96 & - & TBS &48 & - & 96& - & 24&24 &24\\\midrule
     Kernel size  &7 & 2 & 5 & 2 & TBS & 5& 2&7 & 2 &3 &3 &3 \\\midrule
     Stride  &1 & 2 & 1 & 1 & TBS & 1& 2&1 & 2 &1 &1 &1 \\ \bottomrule
    \end{tabular}
}
\vspace{-0.5em}
\end{table}
Motivated by the success of the encoder-decoder structure~\cite{ronneberger2015u} which has demonstrated its efficacy in learning compressed, interpretable, or structured representation of data for denoising, compression, and data completion \cite{eraslan2019single,erhan2010does,tran2017missing, cosentino2020provable}, RT-RCG's DNS engine adopts a search space based on an encoder-decoder based network backbone with searchable blocks to extract and process diverse and patient-specific features from the complex EGM signals. As shown in Table~\ref{tab:backbone} and visualized in Figure \ref{fig:net_struct}, our network starts from a fixed downsample branch and ends in a fixed up-sample branch with the intermediate blocks being set to be searchable for better extracting and processing of the features hidden in the EGM signals. The hypothesis is that such an encoder-decoder structure, i.e., a cascade of convolutional transformations and nonlinearities with a bottleneck dimension, allows the approximation of the underlying data to be manifold as discussed in \citep{cosentino2020provable}.

For the searchable blocks, inspired by the SOTA hardware-friendly search space in~\cite{wu2019fbnet} which searches the kernel size, channel expansion ratio, and group number for each building block, we propose a sequential search space with 14 searchable blocks and 9 candidate operations for each block, including
standard convolutions with a kernel size of 3/5, inverted residual blocks with a kernel size of 3/5, a channel expansion of 1/3/5, and skip connections, which leads to a search space with a total of $9^{14}$ network choices.

\textbf{The Network Search Algorithm.}
We adopt the differentiable NAS (DNAS) algorithm~\cite{liu2018darts} considering its excellent search efficiency. In particular, we formulate the network search as a one-level optimization~\cite{xie2018snas, hu2020dsnas},  by making use of the unbiased gradient estimation~\cite{he2020milenas} to adapt to the complex EGM signals which are diverse for different patients:

\begin{align} 
\vspace{-1.5em}
\small
    \begin{split}
    & \min \limits_{\omega,\alpha} \,\, L_{rec}(\omega, \alpha)+\lambda L^{MAC}_{cost}(\alpha) \label{eq:update_alpha} 
    \end{split} 
\end{align}

\noindent where $\omega$ and $\alpha$ denote the supernet weights and the network architecture parameters, respectively, the latter of which contains the probability of selecting each candidate operation; $L_{rec}$ and $L^{MAC}_{cost}$ denote the ECG-EGM reconstruction loss 
and hardware-cost loss which is determined by the number of multiply-accumulate operations (MACs) in the given DNNs, respectively; and $\lambda$ is a trade-off parameter to balance the resulting reconstruction networks' accuracy and efficiency. In particular, the output of the $l$-th layer $A_l$ in our DNS engine is a weighted sum of all candidate operations:   
\begin{equation}
\label{eqn:arch_loss}
    A_l = \sum_{k=1}^{K} GS(\alpha_{lk}) O_{lk}(A_{l-1})
\end{equation}
\noindent where $K$ is the number of candidate operations, $O_{lk}$ is the $k$-th operation for the $l$-th layer, $\alpha_{lk}$ is the probability of $O_{lk}$, and $GS$ denotes the Gumbel Softmax function~\cite{jang2016categorical} which samples the operations based on the distribution parameterized by $\alpha$. In our DNS, we adopt a soft version of Gumbel Softmax, i.e., we use the output of Gumbel Softmax as the weighted coefficient of $O_{lk}$ with a continuous relaxation during backward pass~\cite{wu2019fbnet} for updating $\alpha$. At the end of the search, we derive the final/searched network by selecting the operation with the highest probability for each searchable block.

\vspace{-0.5em}
\subsection{RT-RCG: The DAS Engine}
\label{sec:DAS}
\vspace{-0.3em}
In this subsection, we introduce the three key components in our proposed DAS engine, i.e., the accelerator template, the search space extracted from the accelerator template, and the search algorithm used to explore the search space.


\vspace{-0.3em}
\subsubsection{The Accelerator Template of Our DAS Engine} \ \\
\label{sec:template}
\vspace{-1.2em}

Our DAS engine leverages a parameterized accelerator template that features a total of $\sim 10^5$ choices for the micro-architecture and dataflow, the latter of which determines how the network is temporally and spatially scheduled to be executed on the micro-architecture, e.g., row stationary, output stationary, weight stationary, etc.

\begin{figure}[t]
    \centering
    \includegraphics[width=0.8\linewidth]{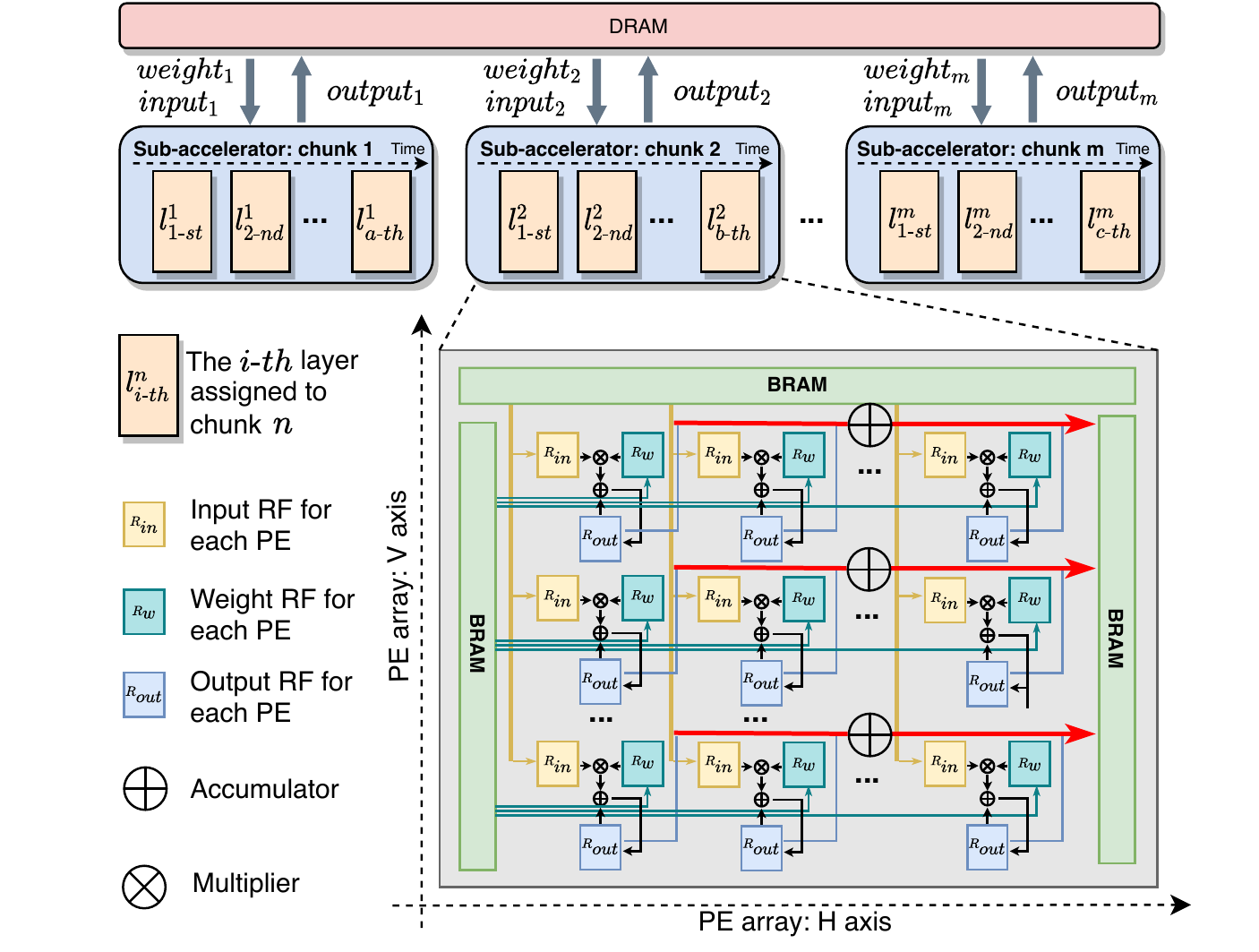}\vspace{-1.5em}
          \caption{An illustration of the parameterized micro-architecture adopted in the DAS engine of our RT-RCG framework.}
    \label{fig:hw_template}
    \vspace{-2em}
\end{figure}
\textbf{The Micro-architecture Overview.} Our DAS engine leverages an accelerator template inspired by a SOTA DNN accelerator~\cite{shen2017ISCA}, which adopts a multi-chunk micro-architecture for maintaining high resource utilization when accelerating DNN layers with different structures (e.g., different sizes of the input/output feature maps and kernel sizes), in order to balance the communication bandwidth and improve the acceleration throughput. Our accelerator template parameterizes the multi-chunk micro-architecture. As illustrated in Figure~\ref{fig:hw_template} later, each chunk of the micro-architecture corresponds to a sub-accelerator, which has hierarchical memories (e.g., on-chip buffer and local register files (RF) ) and processing elements (PEs) characterized by searchable design knobs such as the types of PE interconnections (i.e., Network-on-chip (NoC)), allocated buffer sizes, and the computing scheduling and tiling (i.e., dataflows) to facilitate data reuse and parallelism. Specifically, each sub-accelerator sequentially processes multiple but not necessarily consecutive layers with similar network structures, while different sub-accelerators can be pipelined. 

\textbf{The Sub-accelerator Design.} As shown in Figure~\ref{fig:hw_template}, each sub-accelerator consists of (1) a secondary buffer to facilitate more local data reuse and reduce the higher-cost DRAM accesses and (2) a PE array, where each PE includes a multiply and accumulate (MAC) unit and local register files (RFs) for the inputs, weights, and outputs, respectively. For each sub-accelerator, the dataflow determines the networks' temporal and spatial mapping into the PE array and thus the data movement patterns within different memories/buffers/RF, leading to orders of magnitude difference in the acceleration performance \cite{venkatesanmagnet}. As our accelerator template can parameterize both the micro-architecture and the dataflow (see Section \ref{sec:space}), it enables our DAS engine to search for dedicated micro-architecture and dataflow to match the networks' structure in order to maximize the target hardware performance.

\textbf{Acceleration/Execution.} Here we describe the execution of the network within each sub-accelerator for better understanding. 
In Figure~\ref{fig:hw_template}, if the data within the PEs process different input channels along the H (horizontal) axis of the PE array and different output channels along the V (vertical) axis of the PE array, the weights with different input and output channels will be spatially mapped into all PEs and then stay stationary until all corresponding computations are finished. Meanwhile, the input corresponding to the weights that have been loaded into the PE array will be streamed in via multicast along the H axis and broadcast along the V axis, facilitating various weight reuse. The computed results along the H axis are accumulated while those along the V axis are moved to the output buffer via multicast. In general, the PEs along both axes can process different dimensions of the networks, including the input channels, output channels, feature map height, and feature map width, where the ordering of the subsequent operations and buffer read/write will determine the dataflow and are searchable in RT-ECG.

At any given time point, all sub-accelerators simultaneously process different clusters of the network layers with each sub-accelerator processing data of different input frames, where different layers within each sub-accelerator are executed sequentially, to improve the throughput without the necessity of waiting. This is made possible because (1) sub-accelerators only communicate with the DRAM for fetching/storing the intermediate results and (2) an additional ping-pong buffer is introduced in the DRAM to accommodate simultaneous read/write. In this way, there are no communications needed among the sub-accelerators, leading to a more flexible and modular design. It is then possible to tailor the design of each sub-accelerator to better match the network structure and thus favor the achievable acceleration efficiency.

\vspace{-0.8em}
\subsubsection{The Accelerator Search Space of Our DAS Engine} \ \\
\label{sec:space}
\vspace{-1.2em}

Based on the above accelerator template, we extract the searchable parameters, of which different combinations lead to different accelerators (i.e., micro-architecture and dataflow pairs), to form a generic accelerator space to be used by our DAS engine. The micro-architecture is characterized by the number of memory hierarchies and PEs, the size of each memory hierarchy, the shape and size of the PE array, and the NoC design~\cite{chen2017eyeriss}, and the dataflow is described by both the NoC design and the loop size/order. 
Specifically, we construct a generic accelerator search space as shown in Table~\ref{tab:hw_space} by leveraging the commonly used nested \textit{for-loop} accelerator description~\cite{chen2016eyeriss,parashar2019timeloop,blocking_cnn,zhang2015optimizing,zhao2020icassp} which naturally bridges the accelerator's micro-architectures and dataflows with DNNs' network parameters. Next, we introduce each accelerator parameter listed in Table~\ref{tab:hw_space}: 

\begin{table}[!h]
  \centering
 \vspace{-1.3em}
  \caption{The constructed generic accelerator search space extracted from the accelerator template introduced in Section~\ref{sec:template}, where TBS means ``to be searched'' and the searchable parameters include (1)
 the NoC design, (2) Max \# of PEs, (3) layer assignment,  (4) loop-order and (5) loop-size across different memory hierarchies, i.e., the DRAM, Global Buffer, and PE array.}\vspace{-1em}
    \resizebox{0.5\textwidth}{!}{
        \begin{tabular}{ccc}
        \toprule
        \textbf{Memory Hierarchy} & \textbf{Loop-order} & \textbf{Loop-size} \\
        \midrule
        \textbf{DRAM}   & TBS    & \multicolumn{1}{c}{-} \\
        \textbf{Global Buffer} & TBS    & TBS \\
        \textbf{PE array}    & \multicolumn{1}{c}{-} & TBS \\
        \midrule
        \midrule
        \textbf{NoC design} & \textbf{Max \# of PEs} & \textbf{Layer assignment} \\
        \midrule
        TBS    & TBS    & TBS \\
        \bottomrule
        \end{tabular}%
    }
    
  \label{tab:hw_space}%
\vspace{-1.2em}
\end{table}

\textbf{Loop-order.} The orders of the loops within each memory hierarchy, each of which has a total of $n$ data dimensions. As such, $n$ loops correspond to an $n$-item ordering problem. To be compatible with the proposed network search, where each accelerator parameter should have all possible choices parameterized by the corresponding $\gamma$ vector (see Equation~(\ref{eqn:obj_hw})), we formulate the \textit{loop-order} search as a problem of picking one choice from a total of $n$ options without replacement for $n$ times (e.g., $n=6$ considering the number of data dimensions in DNNs).

\textbf{Loop-size.}
The size of each loop in the \textit{for-loop} description. The product of all loop-sizes associated with each data dimension needs to be equal to the corresponding algorithmic dimension, because the nested loops' size as a whole dictates the total number of execution iterations. Then, intuitively, the possible choices for a certain loop's size are all the choices that the corresponding data dimension can be factorized into.

\textbf{The NoC Design.} The parallel execution patterns of the MAC operations when accelerating DNNs on an accelerator (e.g., those described in Section \ref{sec:template}), which is determined by the PE array style. In this work, we consider three NoC options following the common practice, as inspired by SOTA accelerators~\cite{chen2016eyeriss,zhang2015optimizing,zhao2020icassp}:  

\begin{itemize}
\setlength{\itemsep}{0pt}
    \item Parallelizing the computation over the output partial sums, where the dimensions of output channels, output rows, and output columns are executed in parallel;
    \item Parallelizing the computation over the kernels, where the dimensions of output channels and input channels are executed in parallel;
    \item Parallelizing the computation over both the kernel and output dimensions, where the dimensions of output channels, kernel rows, and output columns are executed in parallel.
\end{itemize}
\textbf{The Maximum Number of PEs.} The maximal number of PEs in the design which can range from 1 to a specified value determined by the area constraint and the trade-off between the storage and computation partition. The PEs will be inter-connected with a pre-designed pattern according to the adopted NoC design, e.g., Figure~\ref{fig:hw_template} gives an example of parallelizing the kernels among the PEs in the NoC across the input and output channel dimensions. In this work, where the latency is the primary objective, the maximum number of PEs is thus set to the hardware platform limit, e.g., the available Digital Signal Processing units (DSPs) in the given FPGAs. If other metrics like energy consumption are prioritized, our proposed framework can automatically search for designs balancing the trade-off between the consumed power and latency.

\textbf{Layer assignment.} The assignment of all the layers to be executed on a fixed number of sub-accelerators, which is set to 10 for this work, unless specified otherwise.\\

\vspace{-1.3em}
With maximum number of PEs fixed and all other parameters above taken into consideration, the space size can explode up to $\sim10^7$. \\

\vspace{-1.3em}
\subsubsection{The Search Algorithm of Our DAS Engine} \ \\
\vspace{-1em}

To efficiently explore our constructed generic accelerator search space, our DAS engine iteratively updates the accelerator design choices in a differentiable manner. In particular, we parameterize the choice of each accelerator design factor with a vector $\gamma$ and learn to update $\gamma$ based on the objective formulated as:

\begin{equation} \label{eqn:obj_hw}
   \gamma^* = \,\, \underset{\gamma}{\min} \,\, \sum_{s=1}^{S} GS(\gamma^s) \, L^{HW}_{cost}(GS(\gamma^1), ..., GS(\gamma^S))\\
\end{equation}

\noindent where $\gamma^s$ defines the probability distribution of the choices for the $s$-th accelerator design parameter, $GS(\gamma^s)$ denotes Gumbel-Softmax sampling~\cite{gumbel1948statistical} of the $s$-th accelerator parameter $\gamma^s$, and $L^{HW}_{cost}(GS(\gamma^1), ..., GS(\gamma^S))$ is the hardware cost of the target network on the sampled accelerator characterized by the $S$ sampled design factors $GS(\gamma^1), ..., GS(\gamma^S)$. 
To be more specific, we apply Gumbel Softmax sampling~\cite{gumbel1948statistical, maddison2014sampling} to sample only one choice $GS(\gamma^s)$ from all the options corresponding to the $s$-th accelerator parameter. Once all the accelerator parameters are sampled, the corresponding accelerator's acceleration cost is estimated using SOTA accelerator performance estimators, where in this work we adopt the performance estimator in~\cite{xu2020autodnnchip} for our prototyped FPGA-based accelerators. After that, we  multiply the resulting acceleration cost by the sampled $GS(\gamma^s)$ and update the $\gamma$ based on the continuous relaxation of Gumbel-Softmax during backward pass~\cite{wu2019fbnet} for gradient estimation. When the gradient-based optimization converges, we derive the final accelerator by selecting the parameter options with the highest probability (i.e., $\gamma_{s}$) for each accelerator parameter. Note that we use the number of MACs as the complexity cost during the network search stage (see $L^{MAC}_{cost}$ in Equation~(\ref{eq:update_alpha})) for better search efficiency, and adopt the estimated accelerator cost $L^{HW}_{cost}$ during the accelerator search stage to better align with the actual acceleration cost. 

\vspace{-0.5em}
\subsection{RT-RCG: The Complexity and Time Cost of The DNS and DAS Engines}
\vspace{0em}
\subsubsection{The Complexity of the DNS and DAS Engines} \ \\
\vspace{-1em}

The algorithm complexity of our DNS engine is tied with that of the supernet training because we adopt the DNAS algorithm as mentioned in Section~\ref{sec:DNS} where the supernet weights and network architecture parameters are updated at the same time. Additionally, picking the final network structure with the highest probability requires an additional complexity of O(k), where k denotes the number of possible operations per block and equals 9 considering our search space defined in Section~\ref{sec:DNS}. On the other hand,
the entire DNS process, including re-training the final picked network, can finish within a GPU hour of 0.5, given the DNS search space size of $9^{14}$ in this work.    

The algorithm complexity of our DAS engine is proportional to that of the Gumbel-Softmax, which is O(n) with n denoting the number of choices for each hardware design parameter. Thanks to the efficient hardware cost estimator~\cite{xu2020autodnnchip}, the entire DAS process only takes about 10 minutes with our space size being $10^7$.

Note that our differentiable search method enables a much more directed and efficient search trajectory. Thus, there is no need to exhaustively evaluate every design choice within the search space, leading to a much shorter search time than that of an exhaustive search. Additionally, the search is terminated when the minimized objectives become stable.

\vspace{-0.5em}
\subsubsection{The Amortized One-Time Search Cost} \ \\
\vspace{-1em}

For a given task, e.g., ECG reconstruction for a specific patient, merely a one-time effort is required to generate the network structure and its accelerator, and thus the search time cost is amortized throughout the implementation. Once the network structure and accelerator design are respectively generated by the DNS and DAS engine, they will be fixed throughout the task. If there are minor changes to the task settings like the patient's heart conditions, the network's parameters (weights) can be fine-tuned with the patient's newly generated heart samples, without the necessity of changing the network structure and accelerator design. The fine-tuning process can be conducted using a standard DNN training procedure on an external computer in a few minutes, considering the setup described in Section~\ref{sec:exp_setup}. Basically, the search only needs to be redone when there are necessary drastic changes, e.g., the change of the entire training dataset. 

\vspace{-0.5em}
\subsubsection{Generalization of the Searched Designs} \ \\
\vspace{-1em}

The searched network structure and its accelerator together with the final fully trained network weights can be generalized to distinct patients' heart samples, if the search and training is conducted on a diverse patient dataset, i.e., 
the searched designs (i.e., the network structure, accelerator design, and trained parameters) are expected to be effective for new patients which are not present in the pre-trained dataset. As such, no additional cost or complexity are incurred for this generalization, as the original search and training process can holistically take the diverse training dataset into consideration. This is validated in Section~\ref{sec:patient_Generalizability}, where the searched networks and accelerator designs consistently perform well on the newly included patients. This generalization capability can be significantly meaningful to real-life applications, where collecting data samples for new patients may not always be possible, and the search cost can thus be amortized across different patients.  
\vspace{-0.5em}
\section{Experiment Results}
\label{sec:exp}
 \vspace{0em}
In this section, we present the evaluation results of our proposed RT-RCG framework. Starting with the introduction to our dataset and experiment setup, we evaluate the effectiveness of the RT-RCG searched networks under various settings, including (1) patient-specific reconstruction (see Section~\ref{sec:single_patient}), (2) reconstruction generalized to a new patient (see Section~\ref{sec:patient_Generalizability}), and (3) robust reconstruction with deficient EGM channels (see Section~\ref{sec:single_channel}). After that, we evaluate RT-RCG's hardware acceleration performance as compared to two SOTA DNN accelerators~\cite{zhang2018dnnbuilder,chaidnn}, one edge platform~\cite{edgegpu}, and a CPU platform, followed by the ablation studies on the initial latency and under a constrained search space. 

\vspace{-0.8em}
\subsection{Experiment Setup}
\label{sec:exp_setup}
\vspace{-0.2em}
\begin{table}[b!]
\vspace{-1.5em}
\centering
\caption{The number of heartbeat samples for each patient and the Patient ID in our clinically collected dataset.}\label{tab1}\vspace{-1em}
\scriptsize
\resizebox{0.95\textwidth}{!}{
    \begin{tabular}{ccccccccccccccc}\toprule
    Patient ID & 1 & 2 & 3 & 4& 5& 6 & 7 & 8 & 9& 10 & 11& 12 &13 &14\\\midrule  \midrule
    Number of heartbeats (N) &4765 &2309 &401 &1752 &3934 &3017 &2593 &6635 &3102 &2326 &5497&1591 &1827 &2917 \\
   
    \bottomrule
    \end{tabular}
}
\vspace{-1em}
\end{table}
\indent\indent \textbf{Clinically Collected Dataset.} To evaluate the effectiveness of the RT-RCG framework, data was collected retrospectively from 14 patients undergoing cardiac ablation for premature ventricular contractions, where both the ECG and EGM signals were recorded simultaneously during the cardiac ablation procedure and each record of the database is composed of:

\begin{itemize}
\item Twelve standard surface ECG channels, namely leads I, II, III, aVR, aVL, aVF, and V1:V6.
\item Five EGM channels measured by electrodes on a catheter placed inside the Coronary Sinus. 
\end{itemize}
Specifically, the data was obtained from patients undergoing cardiac ablation procedures and was retrospectively collected under a protocol approved by an institutional review board at Baylor St. Luke's Medical Center~\cite{BaylorSt65}. During these procedures, the routine is to record both the surface ECG and the EGM signals via the mapping catheter. For each patient, the EGM was obtained from the coronary sinus. By virtue of the procedure, the recordings for each patient are of different lengths and contain a mix of sinus rhythms and diseased heartbeats, providing a diverse dataset to better emulate real-world scenarios while also making it more challenging to achieve high performance reconstruction on this dataset. This also means that the number of heartbeats (i.e., $N$ in Table \ref{tab1}) are different for different patients. In our experiment, the data for each patient was first randomly shuffled and then segmented into halves, with 
the first half of concurrent ECGs and EGMs being used during the search/training step and the second half for testing and performance evaluation. The patient number and corresponding number of heartbeats are 
summarized in Table \ref{tab1}. 


\indent\indent \textbf{Algorithm Experiment Setup.} \underline{Algorithm training setup:} All the DNN training is carried out on a machine with one NVIDIA 2080TI GPU and an AMD EPYC 7742 64-Core power processor. Throughout the training, we use an Adam optimizer with a batch size of 16, a learning rate of 1E-3, and a weight decay factor of 1E-3. During the training, we incorporate the Pearson correlation coefficient between the network output and the ground truth (i.e., corresponding real-measured ECG signals) into the loss function (see Equation (\ref{eq:loss}) in Section~\ref{sec:overview_formulation}).
\underline{Network search setup:} We adopt the one-level optimization as in~\cite{xie2018snas, hu2020dsnas} and a fixed temperature of 1 for the Gumbel Softmax function. We reuse the above training setting for the supernet weights and adopt an Adam optimizer with a constant learning rate of 1E-3 for the architecture parameters. We then derive the operations with the highest probability for each searchable block at the end of the search.
\underline{Algorithm evaluation setup:}  To evaluate the reconstruction efficacy, we calculate the correlation between the reconstructed ECG signals and the real-measured ones on the half of the dataset for testing and performance evaluation. Specifically, we first convert the network output which is in the time-frequency domain to its time domain counterpart using the inverse STFT, and then calculate the Pearson correlation coefficient between the reconstructed signals and the original ECG signals, which are time-series waveforms.

\indent\indent \textbf{Accelerator Experiment Setup.} 
\underline{Accelerator search setup:} Considering the real-time reconstruction goal, we adopt the commonly used Frames Per Second (FPS) metric. However, other metrics can be easily plugged into our RT-RCG framework depending on the specification of the target applications and the user-specified preference.
During the accelerator search process, RT-RCG makes use of a SOTA accelerator performance predictor AutoDNNChip~\cite{xu2020autodnnchip} to obtain a fast and reliable estimation to guide the search towards the optimal solution. 
\underline{Accelerator evaluation setup:} For evaluating FPGA-based accelerators, we adopt a Xilinx ZC706 evaluation board~\cite{zc706} with the same DSP limit as the baselines~\cite{chaidnn,zhang2018dnnbuilder} for a fair comparison. Specifically, we adopt a standard Vivado HLS design flow~\cite{vivado_HLS}, where the FPS is obtained from the HLS synthesis results for our searched accelerators and the baseline ChaiDNN~\cite{chaidnn}. For DNNBuilder~\cite{zhang2018dnnbuilder}, we utilize their open source simulator to obtain its acceleration results. For the CPU baseline, we evaluate the achieved FPS of the networks being executed on an AMD EPYC 7742 64-Core CPU. For the edge platform baseline, we consider a commonly used edge device~\cite{li_2020_halo,siam2018comparative,wofk2019fastdepth}, i.e., the NVIDIA Edge 
GPU Jetson TX2~\cite{edgegpu}, where the networks are compiled using TensorRT~\cite{tensorrt}, a C++ library for high-performance inference on NVIDIA GPUs. Additionally, the device is configured to be in max-N mode to make full use of the available resources following~\cite{wofk2019fastdepth}.

\begin{figure}[t]
    \centering
    \includegraphics[width=0.85\linewidth]{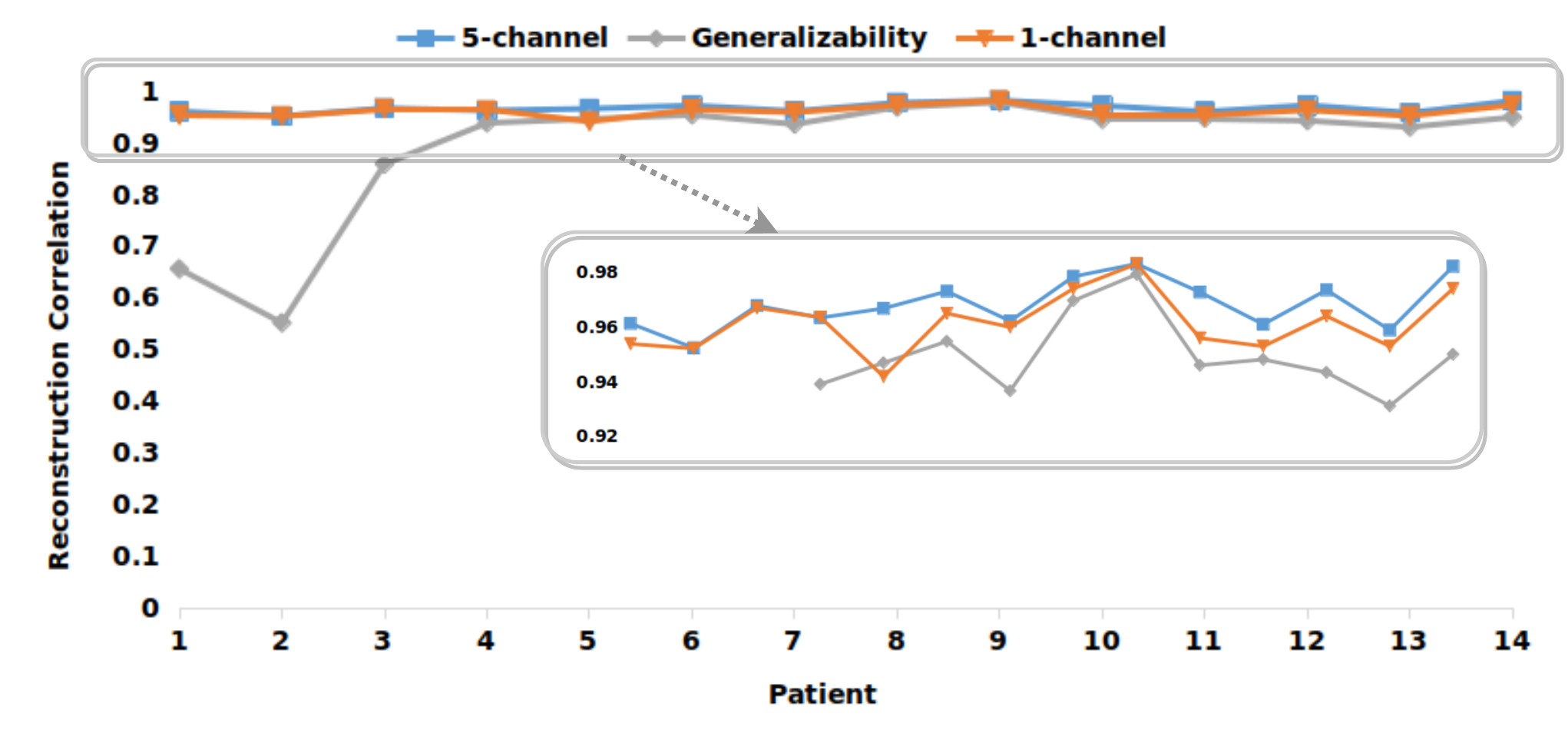}
    \vspace{-2em}
         \caption{ The average Pearson correlation coefficient between RT-RCG's reconstructed and real-measured ECG time-series signals
         across all the 14 patients in our dataset, when considering (1) \textbf{Blue}: patient-specific reconstruction from five channels of EGM (see Section~\ref{sec:single_patient}), (2) \textbf{Grey}: reconstruction generalized to new patients (see Section~\ref{sec:patient_Generalizability}), and (3) \textbf{Orange}: patient-specific reconstruction with merely one EGM channel (see Section~\ref{sec:single_channel}).}
          \vspace{-1.5em}
    \label{fig:network_performance}
\end{figure}
   

\vspace{-1.2em}
\subsection{RT-RCG's Searched Algorithms: Patient-specific Reconstruction}
\label{sec:single_patient}
\vspace{-0.2em}
In this subsection, we evaluate RT-RCG's searched networks in a patient specific setting, where all the search, training, and testing are based on the data collected from the same patients. This is to mimic the case where the pacemakers are customized to each patient. Specifically, for our clinical dataset, which contains sinus and diseased heartbeats of the 14 patients, we equally  split it into two subsets for training and testing, respectively.  

To thoroughly evaluate RT-RCG's searched networks, we consider all of the 14 patients in a patient-specific manner, and plot the resulting correlation (between the constructed ECG and the real-measured ECG signals) in Figure~\ref{fig:network_performance} (the blue curve). We can see that the ECG signals reconstructed by RT-RCG's searched networks are highly correlated with the real-measured ones across all of the 14 patients, as evidenced by the resulting Pearson correlation coefficient value ranging from $0.952\sim0.983$, which is much improved as compared to the correlation value of $0.84$ achieved with the SOTA method~\cite{poree2012surface} using time delay neural networks. This improvement implies that RT-RCG's searched networks can accurately predict ECGs which are close to the corresponding real-measured ones
as compared to the ones reconstructed by the SOTA method in ~\cite{poree2012surface}.

\vspace{-0.8em}
\subsection{RT-RCG's Searched Algorithms: Reconstruction Generalized to New Patients}
\vspace{-0.3em}
\label{sec:patient_Generalizability}
In this subsection, we evaluate the efficacy of our RT-RCG's searched networks when being generalized to new patients. Specifically, the networks are searched and trained based on the data of all patients with one of the patients excluded and then tested on the excluded patient. By doing so, this experiment can evaluate the searched networks' generalization capability to unseen new patients, i.e., how well the networks dedicated to a set of patients can perform when adapted to other patients. As shown in the grey curve in Figure~\ref{fig:network_performance}, the correlation between the reconstructed and real-measured ECG signals is consistently higher than $0.93$, except for Patients 1, 2, and 3, whose heartbeat samples are very distinct from the remaining ones, implying the importance of searching/training the algorithms on diverse patients before being generalized to other patients to ensure the efficacy. Overall, the above experiments indicate the excellent generalization capability of RT-RCG's searched networks. We can expect improved performance if RT-RCG's searched networks are obtained based on more data with diverse ventricular conditions, paving the way for developing ``one-for-all" reconstruction algorithms which can save a large amount of the time and effort needed to collect data for each target patient; this is particularly useful when pre-collecting data for the target patient is not possible.

\begin{figure}[t]
    \centering
    \includegraphics[width=\linewidth]{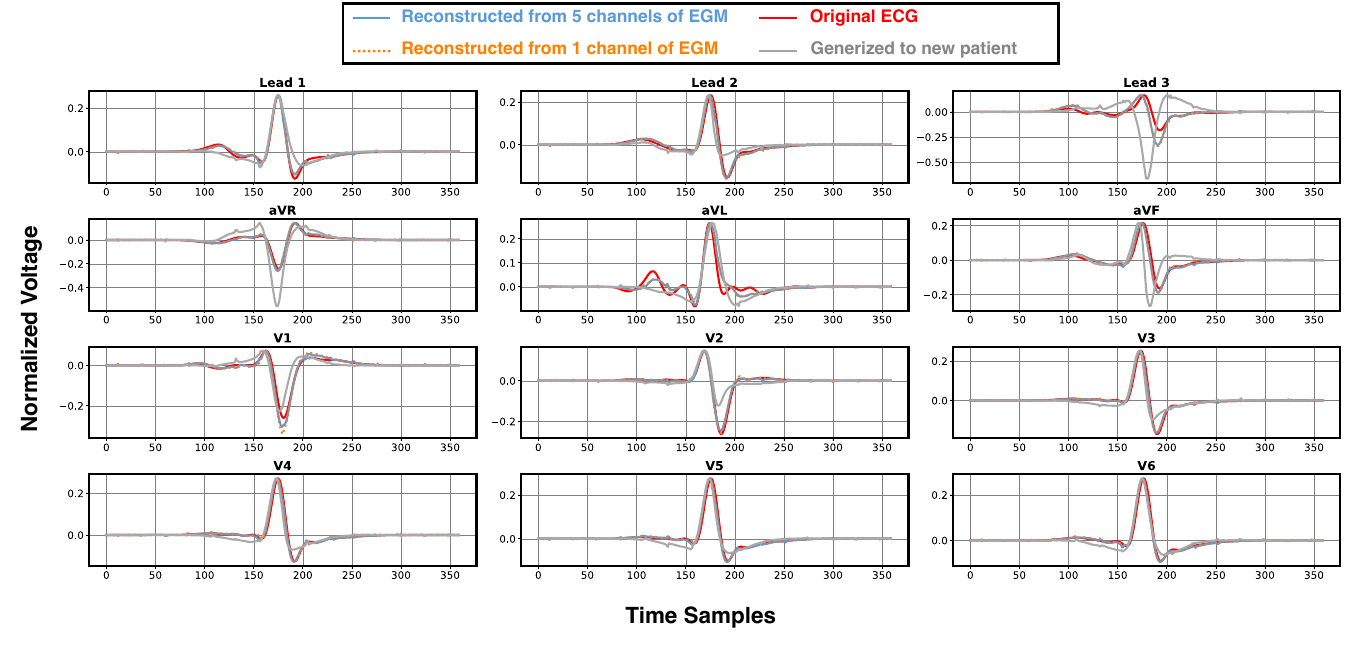}
    \vspace{-3em}
   \caption{ Visualizing the reconstructed ECG signals under different experiment settings together with the corresponding real-measured ones for Patient 4, where the x axis is the time sample and the y axis is the normalized voltage of the waveforms.}
    \label{fig:waveform}\vspace{-1.3em}
\end{figure}

\begin{figure}[t]
  \vspace{-0.5cm}
    \centering
    \includegraphics[width=\linewidth]{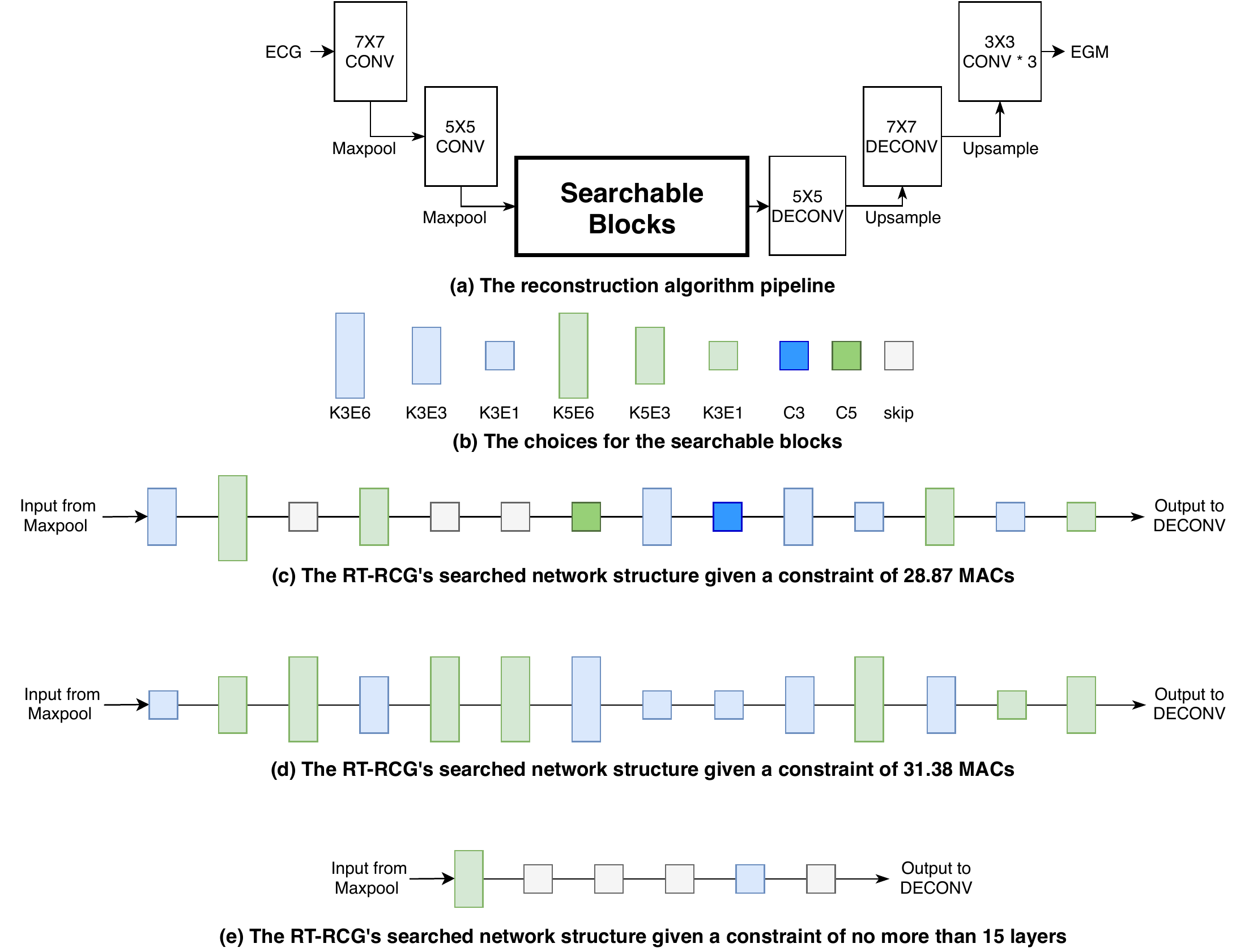}
    \vspace{-2.5em}
    \caption{An illustration of the (a) reconstruction algorithm pipeline, consisting of the fixed earlier blocks, searchable blocks, and fixed later blocks, (b) choices for the searchable blocks following \cite{wu2019fbnet}, and the RT-RCG's searched network structures when given a constraint of (c) 28.87 MACs, (d) 31.38 MACs, and (e) no more than 15 layers. In (b), K\textit{a}E\textit{b} denotes a convolutional building block with a kernel size of \textit{a} and a channel expansion ratio of \textit{b}, and C\textit{a} denotes a standard convolution layer with a kernel size of \textit{a}.}\vspace{-1em}
    \label{fig:net_struct}
\end{figure}

\vspace{-0.5em}
\subsection{RT-RCG's Searched Algorithms: Reconstruction Robustness Under EGM Deficiency}
\vspace{-0.5em}
\label{sec:single_channel}
In practice, pacemakers only utilize 1 - 5 EGM channels and it is an imperative function of pacemakers to work with only one channel of EGM.
Aiming towards practical uses, we thus evaluate our RT-RCG's searched networks under such scenarios, considering the most extreme case where only one out of the five EGM channels is available. Specifically, we search and train the networks based on data with only one EGM channel, and evaluate the correlation between the reconstructed and real-measured ECG signals under the patient-specific setting (similar to Section \ref{sec:single_patient}). While we observe consistent results when picking different EGM channels as the one to be used, we here show the observations when picking the first channel. As shown in the orange curve (i.e., "1-channel") in Figure~\ref{fig:network_performance}, the reconstruction quality under this extreme scenario is surprisingly close to that of the normal setting with all EGM channels on, achieving a correlation ranging from $0.942$ to $0.983$. Furthermore, Figure~\ref{fig:waveform} shows that the reconstructed ECG from only one channel of EGM does not have noticeable degradation when compared with the original ECG signals. This set of experiments demonstrates the excellent robustness of RT-RCG's searched networks in the presence of EGM channel deficiency.



\vspace{-0.5em}
\subsection{RT-RCG's Searched Algorithms: Visualizing the Searched Network and Reconstructed ECG Signals}
\vspace{-0.5em}
\label{sec:single_visual}

To better understand and visualize RT-RCG's searched networks, we here provide a visualization to show RT-RCG's searched network and RT-RCG's reconstructed ECG signals. 
First, as an illustrative example, we visualize the searched network for Patient 4 under a constraint of 28.87M MACs, as illustrated in Figure~\ref{fig:net_struct}(c). In particular, this searched network contains 36 layers excluding the pooling and upsampling layers and a total of 28.87M 
MACs. In addition, the searched networks under different MAC constraints are similar in terms of the kernel size and expansion ratio choices, yet with different preferences in the networks' depth. As shown in Figure~\ref{fig:net_struct}(d), when the number of MACs is increased to 31.38M, the proposed DNS opts to reduce the frequency of skip connections, while the layer structures in terms of kernel sizes and expansion ratios are similar to those under a constraint of 28.87M MACs.
Second, Figure~\ref{fig:waveform} visualizes the reconstructed ECG signals of RT-RCG's searched networks under various settings, when the reconstruction is performed using (1) 5 EGM channels and (2) 1 EGM channel, or generalized to new patients. While we observe consistent results across different patients, here we only show the visualization for Patient 4 for a better illustration. We can see that the reconstructed ECG signals are close to the real-measured ones when the network structure in Figure~\ref{fig:net_struct}(c) is used, with the largest deviation happening when the algorithm is generalized to a new patient, as expected.

\vspace{-0.5em}
\subsection{RT-RCG's Searched Accelerators: Achieved FPS over SOTA DNN Accelerators/Platforms}
\vspace{-0.5em}
\label{sec:fps_hw}
In this subsection, we evaluate RT-RCG's searched accelerators by comparing their achieved FPS with that of (1) two SOTA DNN accelerators (DNNBuilder~\cite{zhang2018dnnbuilder} and ChaiDNN~\cite{chaidnn}), (2) the edge GPU (Jetson TX2~\cite{edgegpu}), and (3) a general DNN deployment platform (an AMD EPYC 7742 64-Core CPU~\cite{2ndGenAM48}) under the same conditions. Specifically, we ensure that the reconstruction algorithm (i.e., the searched network for Patient 4 under 28.87M MACs as shown in Figure~\ref{fig:net_struct}(c)) and the network precision be the same as the baselines'. The comparison results are summarized in Table~\ref{tab:hw_fps}. We can see that RT-RCG's searched accelerator consistently achieves a better FPS than all of the four baselines, based on the same network structure and hardware constraints. Specifically, the RT-RCG searched accelerator improves the achieved FPS, which in turn can be translated to processed heartbeat samples per second, by $1.87\times$, $1.73\times$, $1.22\times$, and $70.90\times$, as compared to the DNNBuilder, ChaiDNN, the edge GPU, and the CPU, respectively. This set of experiments indicates that the integrated DAS engine of RT-RCG is effective and RT-RCG's automatically searched accelerator can even outperform expert designed SOTA DNN accelerators, paving the way for the fast development of reconstruction accelerators.
\begin{table}[!b]
\vspace{-1em}
\centering
\caption{The achieved FPS of the RT-RCG's searched accelerator and the four SOTA DNN accelerators/platforms given the same network (see Figure~\ref{fig:net_struct}(c)), network bit precision, and clock frequency (except for the edge GPU and CPU cases), where  
the number of PEs indicates the peak usage of the processing elements, corresponding to the number of used DSPs for FPGA based accelerators. }\label{tab:hw_fps}\vspace{-1em}
\scriptsize
\resizebox{0.7\textwidth}{!}{
    \begin{tabular}{cccccc}\toprule
    Platform  &Clock frequency &\# of PEs &Bit precision &FPS \\\midrule \midrule
    DNNBuilder~\cite{zhang2018dnnbuilder} &200 MHz &435 &16 & 228\\
    \textbf{RT-RCG} &\textbf{200 MHz} &\textbf{428} &\textbf{16} &\textbf{427} (\boldmath{$1.87\times$}) \\
    \midrule
    ChaiDNN~\cite{chaidnn} &200 MHz &212 &8 & 401\\
    \textbf{RT-RCG} &\textbf{200 MHz} &\textbf{185} &\textbf{8} &\textbf{696} (\boldmath{$1.73\times$}) \\\midrule
    Jetson-TX2~\cite{edgegpu} & 1.3 GHz &/ & 32 & 1190 (183.07 @ 200 MHz) \\
    \textbf{RT-RCG} &\textbf{200 MHz} &\textbf{870} &\textbf{32} &\textbf{229} ($0.19\times$ w/ 1.3 Ghz; \boldmath{$1.22\times$} w/ 200 Mhz) \\\midrule
    CPU~\cite{2ndGenAM48} & 2.25 GHz & / & 32 & 21 (3.23 @ 200 MHz)\\
    \textbf{RT-RCG} &\textbf{200 MHz} &\textbf{870} &\textbf{32} &\textbf{229} ($11.39\times$ w/ 1.3 Ghz; \boldmath{$70.90\times$} w/ 200 Mhz) \\
    \bottomrule
    \end{tabular}
   
}
\vspace{0em}
\end{table}

\begin{table}[b!]
\vspace{-1.5em}
\centering
\caption{The resulting subgroups for the DNNBuilder implementation of the searched network shown in Figure~\ref{fig:net_struct}(c), which is to enable DNNBuilder's feasible implementation of DNNs with over 15 layers. Note that each subgroup assumes one pipeline stage and layers within each subgroup share the same pipeline stage.}\label{tab:net_group}\vspace{-1em}
\scriptsize
\resizebox{0.7\textwidth}{!}{
    \begin{tabular}{cccccccccc}\toprule
    Group ID & 1 & 2 & 3 & 4 & 5 & 6 & 7 & 8 & 9   \\\midrule  \midrule
    Layer ID  &(1) &(2, 3) &(4 $\sim$ 13) &(15 $\sim$ 18) & (19 $\sim$ 26) &(27 $\sim$ 33)& (34) & (35) &(36) \\
    \bottomrule
    \end{tabular}
}
\end{table}

More details regarding the experiment settings for each baseline are described below:

\textbf{DNNBuilder.} For the comparison with the SOTA DNN accelerator named DNNBuilder, we adopt a DSP limit of 450, a 16-bit precision, and a frequency of 200 MHz to be the same as the original setting in DNNBuilder~\cite{zhang2018dnnbuilder}. As the reported DNNBuilder design uses a layer-wise pipeline micro-architecture, it is required to constrain the maximum number of DNN layers to be smaller than 15, for meeting the DRAM access bandwidth constraint, as shown in their open source codes~\cite{zhang2018dnnbuilder}. 
To support RT-RCG's searched networks, which has more than 15 layers, we first divide the network into 9 subgroups with each having some layers from the original network processed sequentially and then execute these subgroups in a pipeline fashion based on the open source design of DNNBuilder. The subgroups are formed to balance the latency among them and thus maximize the achieved throughput of DNNBuilder given the specific DNN structure.
Specifically, the 9 subgroups for the network (see Figure~\ref{fig:net_struct}(c)) are shown in Table~\ref{tab:net_group}.
Note that we also evaluate RT-RCG's searched accelerators over DNNBuilder when constraining RT-RCG's network search space to have networks with smaller than 15 layers as discussed in Section~\ref{sec:15-layer}.

\textbf{ChaiDNN.} We also benchmark RT-RCG's searched accelerator with another SOTA FPGA DNN accelerator named ChaiDNN~\cite{chaidnn}, with its DietChai\_z variant enabled to optimize its performance under more resource constrained scenarios. Specifically, we select its 128-compute-DSP mode which results in a DSP limit of 212 when accelerating the given searched network. 

\textbf{Jetson TX2.} When comparing with the edge GPU Jetson TX2 which is a commonly used IoT device, we set the DSP limit to be 900 (the maximum amount available), so that our implementations have roughly the same power consumption as the edge GPU Jetson TX2. Note that the operating clock frequency of Jetson TX2 is 1.3 GHz, which is far higher than the maximum supported stable frequency of our platform ZC706. We thus scale the Jetson TX2's throughput to that corresponding to a frequency of 200 MHz for a fair comparison as shown in Figure~\ref{fig:net_struct}(c), under which the achieved FPS of the RT-RCG's searched accelerator outperforms the edge GPU by $1.22\times$.


\textbf{CPUs.} Considering that CPUs are currently the mainstream computing platforms, we also evaluate RT-RCG's searched accelerator over an AMD EPYC 7742 64-Core processor given the same network. For a fair comparison, we adopt a DSP limit of 900, which is the maximum available DSP resource on our adopted ZC706 board. Note that the power consumption of the CPU is $\sim225$W, which significantly dwarfs that of the ZC706 board which is $\sim 10$W.

\textbf{Discussion and Implication.} There are several levels of implication from our experiments (including the latency evaluation in Table~\ref{tab:hw_latency}). First, we can see that our proposed RT-RCG indeed can automatically generate (1) reconstruction networks that can provide high-quality reconstruction which outperforms SOTA techniques and has excellent generalization capability and (2) accelerators to run the reconstruction networks that achieve a better acceleration efficiency than diverse SOTA accelerators/platforms, under the same conditions. Second, the performance achieved by RT-RCG shows that it is indeed possible for doctors to remotely monitor the status of pacemakers and patients via reconstructed ECG signals, given the achieved FPS. Specifically, in our case, such real-time monitoring is possible as the achieved FPS (229 $\sim$ 606 FPS in our proposed RT-RCG) is much higher than the required 2 FPS (the highest input rate in our dataset is 2 Hz as it requires at least 0.5s to collect each piece of input).
More importantly, the high FPS achieved is necessary as it implies that real-time intervention is possible, especially when considering that certain cardiac patients, particularly 
patients diagnosed with lethal ventricular arrhythmias,
under which the higher the FPS, the sooner doctors can respond to provide the necessary intervention in life-critical situations. Despite the promising reconstruction efficacy and efficiency achieved by RT-RCG, our effort in this paper is merely a heuristic step towards next-generation pacemakers equipped with real-time monitoring and intervention. In particular, the energy cost of the RT-RCG framework currently implemented on FPGA is still significantly higher than the stringent energy consumption required by the pacemakers. We recognize that applying RT-RCG searched networks and accelerators to real-world pacemakers would require ultra-energy-efficient ASIC implementation, which we leave as one of our most exciting future works. 
\vspace{-0.5em}

\vspace{-0.3em}
\subsection{RT-RCG's Searched Accelerators: Achieved Latency over SOTA DNN Accelerators/Platforms}
\vspace{-0.5em}
\label{sec:latency_hw}
As ECG signals can be used to detect irregular ventricular rhythms which trigger a corresponding alert mechanism~\cite{sukanesh2010gsm}, where the latency from the occurrence of the rhythms to the mechanism being triggered, denoted as start-up latency, can be of great significance to the patients' health and life, the latency of the EGM-ECG conversion contributing a considerable portion of the whole pipeline is thus important. Therefore,
we also evaluate RT-RCG's searched accelerators over SOTA DNN accelerators/platforms in terms of this latency. Note that the achieved start-up latency and FPS have a 
\begin{wraptable}{r}{0.5\textwidth}
\centering
\vspace{-1em}
\caption{The start-up latency and FPS of the RT-RCG generated accelerator given the network generated for Patient 4 (see Section~\ref{sec:single_patient}) under different platforms }\label{tab:hw_latency}\vspace{-1em}
\scriptsize
\resizebox{0.5\textwidth}{!}{
    \begin{tabular}{cccc}\toprule
    Platform & \# of PEs & Start-up latency (ms) &FPS \\\midrule\midrule
    ChaiDNN~\cite{chaidnn} & 212 & 3.01 & 401 \\ \midrule
    RT-RCG & 185 & 3.29 (+9.3\%) & 696 (+73.6\%) \\\midrule
    \textbf{RT-RCG-latency} &\textbf{171} &\textbf{2.39 (+20.6\%)} & \textbf{419 (+4.5\%)}\\
    \bottomrule
    \end{tabular}
}
\vspace{-1.5em}
\end{wraptable}
trade-off relationship.
An advantage of our RT-RCG framework is that users can customize their own desired trade-off given their priority and conditions. 
As shown in Table~\ref{tab:hw_latency}, we provide two searched accelerators of RT-RCG which favor the achieved start-up latency and FPS, respectively, with the former achieving a 27.36\% better start-up latency at a cost of 38.8\% lower FPS. We can see that RT-RCG's automatically searched accelerators achieve a smaller start-up  latency as compared to the baseline under the same hardware constraint, i.e., 20.60\% over the expert-designed accelerator ChaiDNN~\cite{chaidnn}. 
This set of experiments again validates the effectiveness of our RT-RCG framework's DAS engine.

\vspace{-0.5em}
\subsection{RT-RCG's Searched Accelerators: Constrained Networks with <15 Layers}
\vspace{-0.2em}
\label{sec:15-layer}
\begin{table}[t!]
\centering
\caption{The reconstruction accuracy of the searched network with  the constraint of < 15 layers, compared with the searched network in Figure~\ref{fig:net_struct}(c) searched without the layer number constraint. }\label{tab:15-net_acc }\vspace{-1em}
\scriptsize
\resizebox{0.95\textwidth}{!}{
    \begin{tabular}{ccccccccccccccc}\toprule
    Patient ID & 1 & 2 & 3 & 4& 5& 6 & 7 &8 & 9& 10 & 11& 12 &13&14\\\midrule  \midrule
    36-layer-net  &0.9613 &0.9524 &0.9678 &0.9634 &0.9668 &0.9730 &0.9622 &0.9784 &0.9830 &0.9727 &0.9610 &0.9735 &0.9589 &0.9821 \\\midrule
    \textbf{15-layer-net}  &  \textbf{0.9601} &  \textbf{0.9463} &  \textbf{0.9641} &  \textbf{0.9640} &  \textbf{0.9674} &  \textbf{0.9714} &   \textbf{0.9626} &  \textbf{0.9762} &  \textbf{0.9829} &  \textbf{0.9656} &  \textbf{0.9571} &  \textbf{0.9620} &   \textbf{0.9580} &  \textbf{0.9824}\\ \midrule
    \textbf{Improvements from 36-layer}  &  \textbf{-0.0012}& \textbf{-0.0061}& \textbf{-0.0037}& \textbf{0.00057}& \textbf{0.00058}& \textbf{-0.0016}& \textbf{0.00041}& \textbf{-0.0022}& \textbf{-0.00011}& \textbf{-0.0071}& \textbf{-0.0039}& \textbf{-0.011}& \textbf{-0.00088}& \textbf{0.00026}\\
    \bottomrule
    \end{tabular}
}
\vspace{-2.5em}
\end{table}
As mentioned in Section \ref{sec:fps_hw}, the baseline DNNBuilder~\cite{zhang2018dnnbuilder} adopts a layer-wise acceleration micro-architecture, which favors networks with fewer than 15 layers.
To validate the general efficacy of our RT-RCG framework, we here present experiments where we constrain the network search space to \begin{wraptable}{r}{0.5\textwidth}
\centering
\vspace{-1.5em}
\caption{RT-RCG's searched accelerators vs. DNNBuilder, when constraining the networks to have fewer than 15 layers.}\label{tab:15-layer}\vspace{-1em}
\scriptsize
\resizebox{0.5\textwidth}{!}{
    \begin{tabular}{cccc}\toprule
    Platform & \# of PEs &Network MACs (M) & FPS \\\midrule\midrule
    DNNBuilder-36-layer &  435 &28.87&228   \\\midrule
    DNNBuilder-15-layer &  441 &24.23&340 (+49.1\%)   \\\midrule
    \textbf{RT-RCG-36-layer} &\textbf{ 428} & \textbf{28.87}&\textbf{ 427 (+87.3\%)}\\\midrule
    \textbf{RT-RCG-15-layer} &\textbf{ 433} & \textbf{24.23}&\textbf{ 447 (+96.1\%)}\\
    \bottomrule
    \end{tabular}
}
\vspace{-1.5em}
\end{wraptable}ensure that the searched networks have fewer than 15 layers and then compare the acceleration performance of RT-RCG's searched accelerator with that of the DNNBuilder baseline under the patient-specific setting. Specifically, we adaptively adjust $\lambda$ in Equation~(\ref{eq:update_alpha}) when the depth of the derived network surpasses 15 layers by doubling $\lambda$. As shown in Figure~\ref{fig:net_struct}(e), with the number of layers being constrained to 15, the searched network contains only a 16.07\% lower number of MACs as compared to the unconstrained case (see Figure~\ref{fig:net_struct}(c)), while Table~\ref{tab:15-net_acc } indicates that our RT-RCG framework's DNS engine is able to adapt to different constraints while maintaining the networks' performance (i.e., reconstruction quality in terms of the correlation): 0.9601 vs. 0.9613 for Patient 4.
In particular, RT-RCG results in a wider network under this depth constraint in order to maintain the network capacity and thus reconstruction efficacy. 
Meanwhile, as shown in Table~\ref{tab:15-layer}
we can see that (1) DNNBuilder's achieved FPS is improved by 49.1\% as compared to the unconstrained case presented in Section~\ref{sec:fps_hw}, which has a 36-layer network, under the same DSP constraint, and (2) RT-RCG's automatically searched accelerator again outperforms the expert designed accelerator DNNBuilder with a 23.94\%  higher FPS. This set of experiments together with the ones in Section~\ref{sec:fps_hw} validates the general effectiveness of our RT-RCG framework across different network search spaces and accelerated networks.

\section{Conclusion}

The costly and time-consuming hospital visits required for patients with implanted pacemakers and the recent advances in the IoT technologies have motivated an increasing need for remote monitoring of pacemakers to reduce hospital visit costs and to provide continuous monitoring and potential real-time intervention which can be life-critical under some irregular and infrequent ventricular rhythms. However, the signals provided by pacemakers and the ones doctors use for diagnosis during in-person clinical visits are different, with the former being EGM signals and the latter being ECG signals, calling for high-quality and real-time ECG reconstruction from the recorded 
EGM signals. To this end, we propose, design, and validate a first-of-its-kind framework dubbed RT-RCG, which can automatically search for (1) efficient DNN structures and then (2) corresponding hardware accelerators to implement the ECG-EGM reconstruction process, respectively, tackling both the reconstruction efficacy and efficiency.  
Specifically, RT-RCG integrates a new DNN search space tailored for required ECG-EGM reconstruction to enable automated search for DNNs that conduct ECG reconstruction with much improved quality over SOTA solutions, and incorporates a differentiable acceleration search engine that can automatically generate optimal accelerators to accelerate the resulting DNNs from the previous step. Extensive experiments and ablation studies under various settings consistently validate the effectiveness and advantages of the proposed RT-RCG at leading to higher reconstruction quality and better reconstruction efficiency as compared to SOTA reconstruction algorithms and DNN accelerators. Our RT-RCG has made the first heuristic step towards automated generation of  ECG-EGM reconstruction DNNs along with the matched accelerators, which enable real-time critical intervention in instant response to irregular and infrequent ventricular rhythms that require timely treatment, paving the way for more pervasive remote monitoring of the pacemakers via ECG-EGM reconstruction.

\bibliographystyle{ACM-Reference-Format}
\bibliography{acmart}

\appendix


\end{document}